\documentclass[pra, reprint, twocolumn, superscriptaddress, amsmath, amssymb, aps, ,floatfix]{revtex4-2}

\usepackage[english]{babel}
\usepackage{graphicx}
\graphicspath{{./Images/},{./ImagesAppendix/},{./images/},{./imagesAppendix/}}

\usepackage{amsmath, amssymb, amsfonts}
\usepackage{physics}
\usepackage{bm}
\usepackage{braket}
\usepackage{array}
\usepackage{multirow}
\usepackage{float}
\usepackage{commath}
\usepackage{verbatim}
\usepackage{color}
\usepackage{enumitem}
\usepackage{array}   
\usepackage{dsfont}
\usepackage{mathtools}
\usepackage{soul}
\usepackage[colorlinks=true, citecolor=darkBlue, linkcolor=darkBlue, urlcolor=blue]{hyperref}
\usepackage{amssymb}
\usepackage{tabularx}

\definecolor{darkBlue}{rgb}{0.08, 0.13, 0.4}

\usepackage{tikz}
\usetikzlibrary{positioning, arrows.meta}
\usepackage{pgfplots}

\definecolor{THc}{rgb}{0.9,0.3,0.2}

\newcommand{\canc}[1]{}

\begin{document}

\title{Floquet-engineered fidelity revivals in the PXP model}
\author{F. Perciavalle}
\affiliation{Dipartimento di Fisica, Universit\`a della Calabria, 87036 Arcavacata di Rende (CS), Italy}
\affiliation{INFN--Gruppo collegato di Cosenza}

\author{F. Plastina}
\affiliation{Dipartimento di Fisica, Universit\`a della Calabria, 87036 Arcavacata di Rende (CS), Italy}
\affiliation{INFN--Gruppo collegato di Cosenza}

\author{N. Lo Gullo}
\affiliation{Dipartimento di Fisica, Universit\`a della Calabria, 87036 Arcavacata di Rende (CS), Italy}
\affiliation{INFN--Gruppo collegato di Cosenza}

\date{\today}

\begin{abstract}
We explore the dynamics of the PXP model when subjected to a periodic drive, and unveil the mechanism through which the interplay between spectral properties and initial states governs the emergence of dynamical revivals and their evolution across the space of driving parameters. For Néel-ordered initial states, revivals follow well-defined trajectories in the parameter space of the driving, primarily determined by a dominant quasi-energy spacing in the Floquet spectrum. Initial states interpolating between Néel and fully polarized configurations exhibit hybrid dynamics, which can be controlled by tuning their overlap with Floquet eigenstates via the driving parameters. This control also allows steering different routes for avoiding Floquet thermalization, showing how both initial state choice and driving protocol shape long-lived dynamics in this driven quantum many-body systems.

\end{abstract}

\maketitle

\section{Introduction}
In recent years, controlling the dynamics of quantum matter has emerged as a highly active research field, driven by the advent of versatile and programmable quantum platforms~\cite{georgescu2014quantum, daley2022practical,browaeys2020many, aspuruguzik2012photonic, altman2012quantum, amico2021roadmap, polo2024perspective, morsch2025quantum, polo2025persistent}. Within this broad landscape, the study of thermalization in isolated quantum systems has attracted significant attention from both theoretical and experimental points of view~\cite{dalessio2014long, rigol2008thermalization, srednicki1994chaos, bernien2017probing, turner2018quantum, pal2010many, capizzi2025hydrodynamics, kaufman2016quantum, smith2016many}. A key concept in understanding thermalization is the Eigenstate Thermalization Hypothesis (ETH)~\cite{dalessio2014long, rigol2008thermalization, srednicki1994chaos}, which posits that individual energy eigenstates of a generic interacting quantum system encode thermal behavior, such that expectation values of local observables approach their thermal averages in the long-time limit. While ETH provides a robust framework for typical systems, certain non-ergodic phenomena escape its description. In particular, quantum systems such as the PXP model~\cite{turner2018quantum, lesanovsky2012interacting, ho2019periodic, serbyn2021quantum, hudomal2022driving} host quantum many-body scars (QMBS)~\cite{serbyn2021quantum, chandran2023quantum, choi2019emergent, pappalardi2025theory, desaules2022extensive, kerschbaumer2025quantum, daniel2023bridging, pizzi2025genuine, perciavalle2025local, banerjee2025quantum, surace2021exact}, which are atypical eigenstates characterized by anomalously low entanglement, allowing for persistent coherent dynamics and periodic revivals from specific initial states, despite the surrounding spectrum being largely thermal.

In addition, periodic driving provides a powerful tool to further control and engineer the dynamics of quantum many-body systems. Such systems, in which the Hamiltonian is time-periodic, can exhibit effective long-time behavior that differs significantly from static systems. Floquet theory provides a framework to understand these dynamics by describing the evolution in terms of stroboscopic steps at integer multiples of the driving period, effectively defining a Floquet Hamiltonian that governs the system at discrete times~\cite{eckardt2015high, santoro2019introduction, russomanno2014periodic, domanti2024floquet, geier2021floquet, sato2025floquet, holthaus2015floquet}. This approach enables access to novel dynamical regimes, including engineered non-ergodic behavior, subharmonic responses and time-crystals, controlled revivals, which can be tuned by varying the drive parameters such as amplitude and frequency~\cite{bluvstein2021controlling,hudomal2022driving,dutta2025prethermalization,park2023subharmonic,dutta2025controlling, tang2025discrete, maskara2021discrete, deng2023using, luo2025discrete}.

\begin{figure}[!t]
\centering
\includegraphics[width=0.9\linewidth]{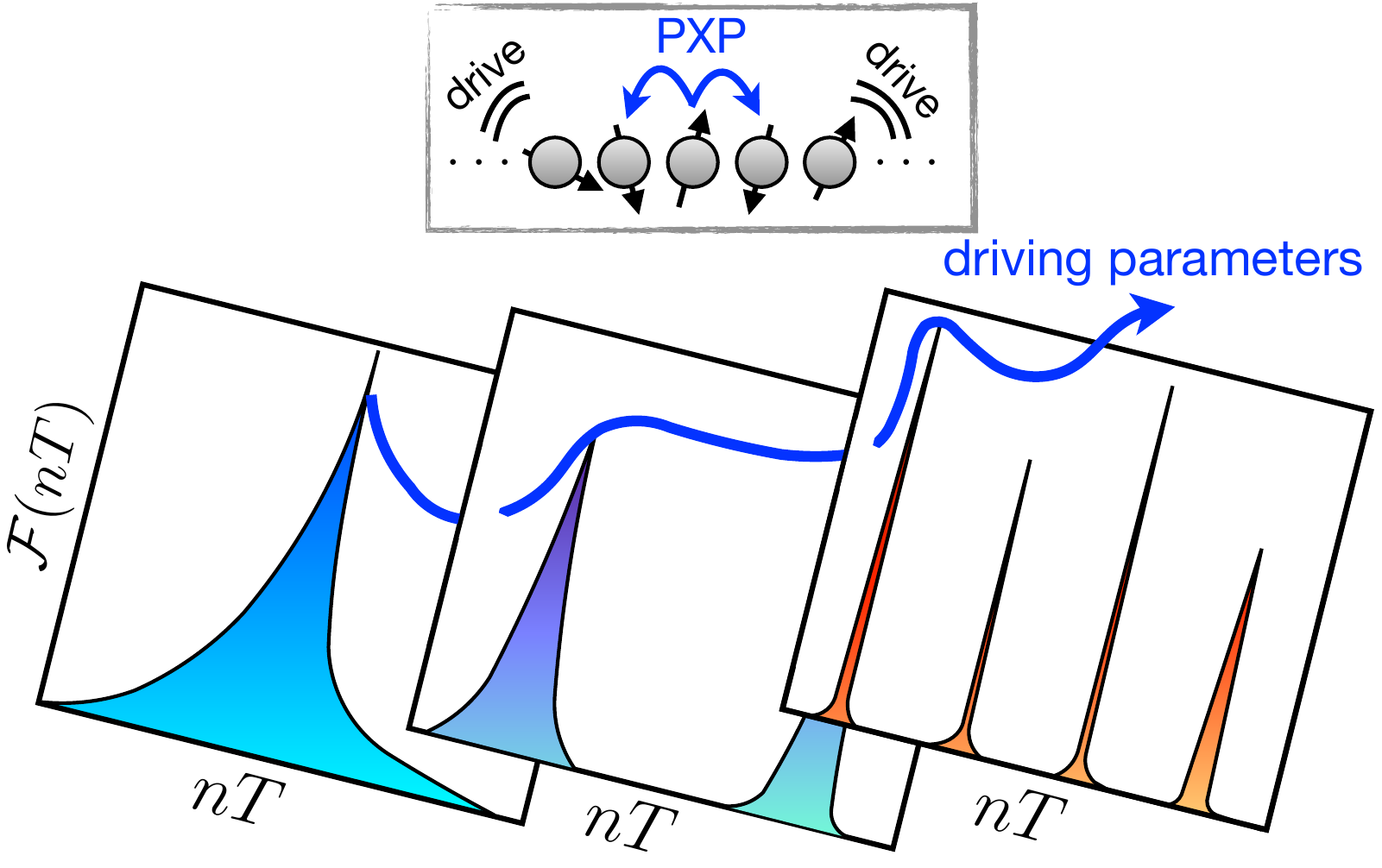}
\caption{We consider a periodically driven quantum PXP model, schematically illustrated in the box at the top of the figure. When the system is initialized in the Néel state, the driving parameters control the revival dynamics, which follows a well-defined pathway. For other initial states, tuning the same driving parameters can give rise to alternative pathways, interpolating between qualitatively different dynamical regimes. Consequently, the system can access distinct dynamical regimes, as indicated by the blue arrow traversing different regions of the diagram.}
\label{fig:sketch}
\end{figure}
In this paper, we investigate the mechanisms behind the emergence of revivals in the driven PXP model and their connection to the Floquet spectrum. We analyze the dynamics starting from different initial states and track the evolution of revivals in the driving parameter space, as illustrated in Fig.~\ref{fig:sketch}, showing how their occurrence depends on some driving parameters, such as amplitude and frequency. For the Néel state, we identify a well-defined revival mechanism, whereas for initial states interpolating between the Néel and fully polarized states the revival behavior can be continuously tuned via the driving parameters. More interestingly, we show that the driving parameters can also be used to tune between different dynamical regimes, namely, erratic and Néel-like, while keeping the initial state fixed. Distinct dynamical behaviors arise from differences in the projection of the initial state onto the Floquet states, which can be controlled through the driving protocol. Moreover, we investigate the tendency of the system to thermalize under periodic driving within the framework of Floquet thermalization~\cite{haldar2018onset,staszewski2025krylov,seetharam2018absence,hou2025floquet,giudici2024unraveling}, which predicts that generic driven systems absorb energy and eventually approach an effective infinite-temperature state. This state may be reached on average (weak thermalization) or at specific times (strong thermalization). We find that the dynamics of the polarized state are at least consistent with weak thermalization, whereas the Néel state does not exhibit thermalizing behavior. For initial states interpolating between these two limits, the tunable overlap structure gives rise to distinct mechanisms for avoiding thermalization, which can be controlled through the driving parameters.

The remainder of the paper is organized as follows. In Sec.~\ref{sec:pxp_floquet}, we introduce the driven PXP model and provide an overview of Floquet theory. In Sec.~\ref{sec:revival_mechanism}, we analyze the revival mechanism in the driving parameter space for a system initialized in the Néel state, while in Sec.~\ref{sec:spectral_origin}, we investigate its spectral origin. In Sec.~\ref{fig:theta}, we analyze the revival mechanism for initial states interpolating between the Néel and fully polarized states, and in Sec.~\ref{sec:thermalization}, we discuss their thermalization properties. Finally, in Sec.~\ref{sec:conclusions}, we present our conclusions and outline perspectives arising from this work.
\section{Driven PXP model and Floquet theory}
\label{sec:pxp_floquet}
We consider a quantum spin chain composed of $L$ qubits described by a periodically driven PXP model with open boundary conditions (OBC), whose Hamiltonian reads
\begin{align}
&\hat{H}(t)=\hat{H}_{\rm PXP} -V(t)\hat{N},\label{eq:hamiltonian}\\
&\hat{H}_{\rm PXP}=\frac{\Omega}{2}\large(\hat{X}_1 \hat{P}_2 + \sum_{j=2}^{L-1}\hat{P}_{j-1}\hat{X}_j \hat{P}_{j+1} + \hat{P}_{L-1}\hat{X}_L\large), \nonumber\\
&V(t)\hat{N}=h \sin(\omega_d t)\sum_{j=1}^L \hat{N}_j=h \sin(\omega_d t)\sum_{j=1}^L \frac{1}{2}(\hat{\mathds{1}}_j + \hat{Z}_j),\nonumber
\end{align} 
where $\hat{Z}_j = \ket{1}_j\bra{1} - \ket{0}_j\bra{0}$ is the $z$-Pauli matrix, $\hat{X}_j = \ket{1}_j\bra{0} + \ket{0}_j\bra{1}$ is the $x$-Pauli matrix, $\hat{P}_j = \frac{1}{2}(\hat{\mathds{1}}_j - \hat{Z}_j)$ is the projector onto the state $\ket{0}_j$, while $\hat{N}_j = \ket{1}_j\bra{1}$. In the following, we denote the remaining Pauli matrix as $\hat{Y}_j$. The parameters $h$ and $\omega_d$ are the driving amplitude and frequency, respectively, while $L$ denotes the system size. Throughout this work, we set $\hbar=1$, and measure energy (and time) in units of $\Omega$ (and $\Omega^{-1}$).

The PXP model naturally arises in systems of interacting Rydberg atoms operating in the blockade regime~\cite{turner2018quantum, bernien2017probing, lesanovsky2012interacting, ho2019periodic, serbyn2021quantum, liang2025observation, surace2021exact}. The periodic driving term can be experimentally implemented through a time-dependent detuning~\cite{bluvstein2021controlling,hudomal2022driving,dutta2025prethermalization,park2023subharmonic,dutta2025controlling}. In the absence of driving, the PXP model exhibits a rich dynamical phenomenology characterized by a sensitive dependence on the choice of the initial state. When initialized in the fully polarized state $\ket{\boldsymbol{0}}=\ket{0000\ldots}$, the system rapidly explores the available Hilbert space and displays ergodic behavior consistent with the ETH~\cite{turner2018quantum}. By contrast, initialization in the Nèel state $\ket{\mathbb{Z}_2}=\ket{1010\ldots}$ leads to a pronounced violation of ETH and to long-lived quantum revivals~\cite{turner2018quantum}. These revivals originate from the overlap of the Nèel state with a set of atypical many-body eigenstates, known as QMBS, which exhibit anomalously low entanglement within an otherwise thermal spectrum and are responsible for coherent oscillatory dynamics. More recently, it has been shown that the PXP model hosts additional families of non-ergodic initial states beyond the polarized and Nèel configurations. These have been defined local reminiscent states, as they preserve a substantial memory of their initial local structure over long times~\cite{perciavalle2025local}, further highlighting the diversity of dynamical behaviors supported by the model.

The dynamics of the model is restricted to a subspace in which no two adjacent excitations are allowed, a condition known as Rydberg blockade. In fact, the Hamiltonian commutes with the operator $\hat{\mathcal{P}}=\prod_j(1 - \hat{N}_j \hat{N}_{j+1})$~\cite{omiya2023fractionalization}, which acts as the identity on states within the blockaded subspace and annihilates states outside it. Consequently, any initial state in this subspace evolves entirely within it. For such states, the effective Hilbert space dimension is reduced from $2^L$ to a Fibonacci-scaling subspace of size $F_{L+2}$, where $F_n$ is the $n$-th Fibonacci number~\cite{bernien2017probing,turner2018quantum}. The drive considered in Eq.~\eqref{eq:hamiltonian} also commutes with $\hat{\mathcal{P}}$, so the blockade constraint remains valid in the driven model.

\begin{figure*}[!t]
\centering
\includegraphics[width=\linewidth]{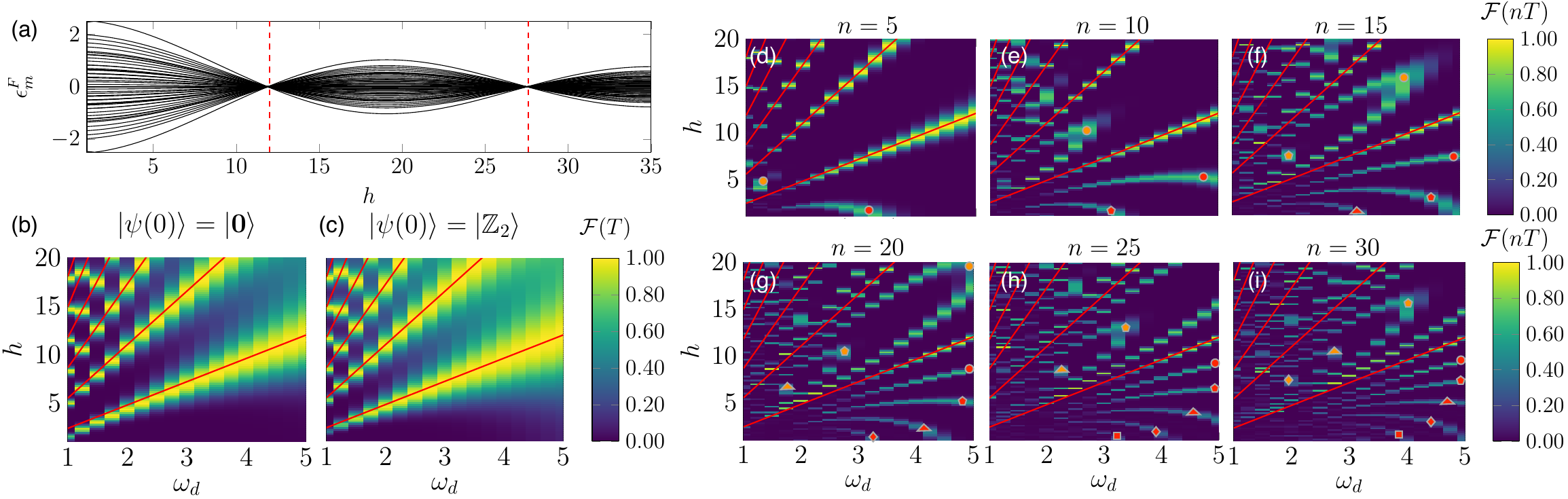}
\caption{Panel (a): Floquet spectrum as a function of the driving amplitude $h$ with $\omega_d=5$ and $L=8$ fixed. The vertical dashed red lines indicate the first two zeros of the $J_0(h/\omega_d)$ Bessel function: $h/\omega_d \approx 2.4048,\, 5.5201$. Panel (b): one-period fidelity $\mathcal{F}(T)$ of the polarized $\ket{\boldsymbol{0}}$ (panel (b)) and the Nèel $\ket{\mathbb{Z}_2}$ (panel (c)) initial states as a function of both $h$ and $\omega_d$, with $L=12$. The red lines correspond to the zeros of $J_0(h/\omega_d)$. Panels (d),(e),(f),(g),(h),(i) report the fidelity between initial state (the system is initialized in the Nèel state) and time-evolved state at different stroboscopic times, labeled by $n$ on the top of each panel, with $L=12$. The evolution mechanism has the structure of a wave propagating in specific spatial regions identified by the zeros of the Bessel function. The spatial plane is fictitious and it is given by the driving parameters $h$ and $\omega_d$. We follow the crests of the evolving waves by the different red and orange symbols. Red symbols follow crests in the $h/\omega_d<2.4048$ region and orange symbols in the $2.4048<h/\omega_d<5.5201$.}
\label{fig:memory1}
\end{figure*}
In general, the inclusion of a periodic drive further enriches the dynamical landscape of the PXP model, giving rise to phenomena such as subharmonic responses and enhanced fidelity revivals in specific parameter regimes~\cite{bluvstein2021controlling,hudomal2022driving,dutta2025prethermalization,park2023subharmonic,dutta2025controlling,mukherjee2020collapse}. In general, a periodically driven quantum system with Hamiltonian satisfying $\hat{H}(t)=\hat{H}(t+T)$, where $T=2\pi/\omega_d$, can be analyzed within the framework of Floquet theory~\cite{holthaus2015floquet,santoro2019introduction,haldar2012dynamical,eckardt2015high}. The time evolution operator, $\hat{\mathcal{U}}(t,0) = \mathcal{T}\exp\left(-i\int_{0}^{t} dt'\,\hat{H}(t')\right)$, can be expressed in terms of repeated applications of the single-period evolution operator, $\hat{\mathcal{U}}(t+nT , 0) = \hat{\mathcal{U}}(t , 0) \big[ \hat{\mathcal{U}}(T , 0) \big]^{n}$. At stroboscopic times $t=nT$, the dynamics is fully governed by the Floquet operator $\hat{\mathcal{U}}(T,0)$. A generic time-evolved state at stroboscopic times can be written as
\begin{equation}
\ket{\psi(nT)}=\sum_m c_m e^{-i\epsilon_m^F nT}\ket{\epsilon_m^F}, \quad
c_m=\braket{\psi(0)|\epsilon_m^F},
\end{equation}
where ${\ket{\epsilon_m^F}}$ denote the Floquet eigenstates of the single-period evolution operator, and $\lambda_m^F=e^{-i\epsilon_m^F T}$ are the corresponding eigenvalues. The quantities $\epsilon_m^F$ are referred to as quasi-energies and are defined modulo $\omega_d=2\pi/T$, since $\exp[-i(\epsilon_m^F+\omega_d)T]=\exp(-i\epsilon_m^F T)$. As a consequence, quasi-energies are restricted to the Floquet Brillouin zone (FBZ), $\epsilon_m^F\in(-\omega_d/2,\omega_d/2]$. The structure of the Floquet spectrum therefore plays a central role in determining the long-time dynamics of the system and naturally motivates the definition of a Floquet Hamiltonian $\hat{H}^F$, implicitly defined through $\hat{\mathcal{U}}(T,0)=\exp(-i\hat{H}^F T)$.

Periodic driving can thus be viewed as an effective tool to engineer renormalized Hamiltonians and access dynamical regimes that are difficult to realize in static settings. In particular, for the driven PXP model it has been shown that, in suitable perturbative regimes, the effective Floquet Hamiltonian can be expressed as a mixture of PXP and PYP terms~\cite{dutta2025prethermalization}. The relative importance of these contributions is controlled by the ratio $h/\omega_d$, which sets the characteristic energy scale of the Floquet quasi-energies. At leading order in a high-frequency expansion, the effective Hamiltonian scales as $J_0(h/\omega_d)$~\cite{dutta2025prethermalization}, where $J_0$ denotes the Bessel function of the first kind of order zero.

In the following, we are interested in shedding light on the mechanism that governs the appearance of quantum revivals in the driven PXP model, by tracking the trajectory of the revivals in the driving parameter space, as sketched in Fig.~\ref{fig:sketch}. Specifically, we aim to understand the factors and conditions that regulate their emergence.
\section{Revival mechanism in the driven PXP model}
\label{sec:revival_mechanism}
In this section, we perform a general analysis of the dynamics at stroboscopic times, where the evolution is fully determined by the Floquet spectrum of the system. This approach allows us to identify the effects of the driving and to pinpoint the spectral features responsible for the revival behavior. Consequently, the first step is a careful analysis of the Floquet spectrum for different values of the driving parameters. According to Ref.~\cite{dutta2025prethermalization}, in the high-amplitude and high-frequency regime, the Floquet Hamiltonian can be effectively described as a combination of the PXP and PYP models, modulated by the factor $J_0(h/\omega_d)$, except at the zeros of this function, which are $h/\omega_d \approx 2.4048,\,5.5201,\,\ldots$.

In Fig.~\ref{fig:memory1}(a), we show the Floquet quasi-energies as a function of the driving amplitude $h$, for fixed $\omega_d=5$ and $L=8$. The spectrum exhibits a distinctive candy-like structure, with a Floquet bandwidth, defined as the difference between the highest and lowest quasi-energy, that approximately follows $|J_0(h/\omega_d)|$, in agreement with perturbative predictions. For driving amplitudes smaller than the driving frequency, the quasi-energies spread over the full FBZ, delimited by $-\omega_d/2$ and $\omega_d/2$. In the vicinity of the zeros of $J_0(h/\omega_d)$, the Floquet spectrum narrows significantly, although it retains a finite bandwidth. We refer to this region as Floquet spectrum narrowing (FSN). As we will discuss, this FSN region plays a central role in controlling the revival mechanism and understanding the emergence of coherent dynamics in the driven system.

\begin{figure*}[!t]
\centering
\includegraphics[width=\linewidth]{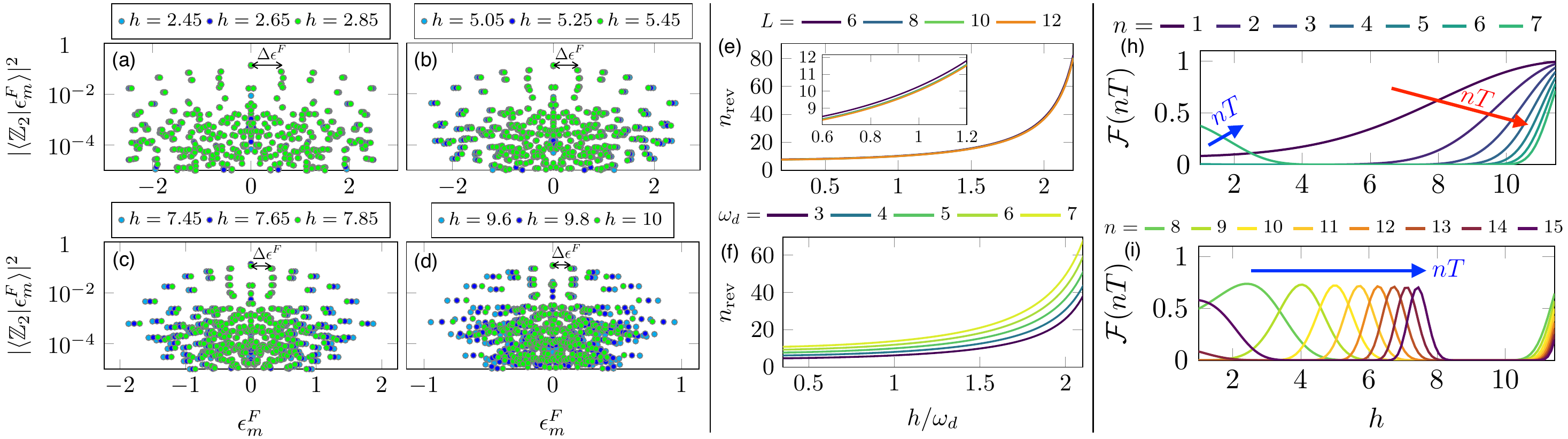}
\caption{Spectral properties and dynamics of the $\ket{\mathbb{Z}_2}$ state in the regions of $h/\omega_d<2.4048$, i.e. the first zero of $J_0(h/\omega_d)$.  
Panels (a)-(d): deformation of the overlap between the Nèel state and the Floquet states for different values of the driving amplitude $h$, with $\omega_d=5$ and $L=12$. The four panels are associated to three specific regions of the driving amplitude $h$. Panels (e) and (f): revival time $n_{\rm rev}$ estimated from the Floquet spectrum as $n_{\rm rev}=\frac{T_{\rm rev}}{T}$ with $T_{\rm rev}=2\pi/\Delta\epsilon^F$, with $\Delta\epsilon^F$ extracted from the overlap between Nèel state and Floquet states, as a function of the driving amplitude $h$, normalized by the driving frequency $\omega_d$. In panel (e), we report it for different values of the size of the system $L$, with $\omega_d=5$. The inset contains the same object but zoomed in a specific region of $h$. Panel (f) reports the same quantity but for different values of $\omega_d$ and with $L=10$ fixed. Panels (h),(i) report the $n$-period fidelity $\mathcal{F}(nT)$ of the system initialized in the Nèel state as a function of $h$: panel (h) reports early times $n\leq 7$ and panel (i) reports later times $n\in [8,15]$. The red and blue arrows guide the eye along the time evolution of the revivals during the decay and revival phases, respectively. We consider $L=12$ and $\omega_d=5$.}
\label{fig:spectral_fig}
\end{figure*}
To explore this mechanism, we follow the stroboscopic fidelity
\begin{equation}
\mathcal{F}(nT)=|\braket{\psi(nT)|\psi(0)}|^2,
\end{equation}
which quantifies the similarity between the initial state and the time-evolved state at stroboscopic times. It serves as a powerful tool to probe the periodic dynamics of the system. By analyzing the time evolution at stroboscopic times, we can capture the key features that lead to the emergence of revivals and gain insight into the conditions that govern their appearance. 
Noting that $\hat{\mathcal{U}}(nT,0)=\big[\hat{\mathcal{U}}(T,0)\big]^n$, the stroboscopic fidelity satisfies $|\braket{\psi((n+k)T)|\psi(kT)}|^2 = |\braket{\psi(nT)|\psi(0)}|^2 = \mathcal{F}(nT)$.
Therefore, the stroboscopic fidelity generally provides a measure of memory degradation over time intervals separated by $nT$, capturing how well the system retains its initial configuration across stroboscopic periods. 

Figs.~\ref{fig:memory1}(b),(c) show $\mathcal{F}(T)$, which quantifies the memory loss over a single period, for both the polarized $\ket{\boldsymbol{0}}$ and the Nèel $\ket{\mathbb{Z}_2}$ initial states, with the system size fixed at $L=12$. $\mathcal{F}(T)$ is particularly large near the FSN region, due to the shrinkage of the Floquet bandwidth and the corresponding elongation of the stroboscopic timescales of the dynamics. Away from this region, the one-period memory loss can be tuned by varying both the driving parameters and the choice of initial state. For example, for $h/\omega_d < 2.4048$, $\mathcal{F}(T)$ is close to zero for both initial states, indicating almost complete loss of memory. In contrast, at high driving amplitudes and frequencies, $\mathcal{F}(T)$ increases significantly, with the Nèel state exhibiting better memory preservation over a single period. In summary, this result, which mainly derives from Fig.~\ref{fig:memory1}(a), provides a clear guideline for achieving the desired memory behavior over a single period: strong preservation in the FSN regions, negligible preservation for $h/\omega_d < 2.4048$, and partial preservation, particularly for the Néel state at high driving amplitude and frequency.

We now focus on the revival mechanism, quantified by the $n$-period fidelity $\mathcal{F}(nT)$ when the system is initialized in the Nèel state. This quantity not only measures the recurrence of the initial state at stroboscopic times, but also captures the memory degradation over intervals of $nT$, since it quantifies the similarity between $\ket{\psi(0)}$ and $\ket{\psi(nT)}$, as well as between $\ket{\psi(nT)}$ and $\ket{\psi(2nT)}$, and so forth. Figs.~\ref{fig:memory1}(d)-(i) show the evolution of $\mathcal{F}(nT)$ at selected stroboscopic times. From these plots, we identify the emergence of a mechanism governing the appearance of revivals in the system. We first observe that in the FSN regions, indicated by red lines, the dynamics is significantly slowed down since the Floquet bandwidth is strongly compressed. Far from the FSN regions, revivals initially appear at small values of the driving amplitude $h$ and then progressively move through parameter space as time evolves. This propagation qualitatively resembles a wave traveling in a fictitious space defined by $(h,\omega_d)$, with the FSN regions effectively acting as barriers that the wave cannot cross on short timescales. At sufficiently long times, the slowing down induced by the FSN regions is overcome, and the system develops nontrivial dynamics even in these parameter regimes. 

In the figure, we focus on the regions $h/\omega_d < 2.4048$ and $2.4048 < h/\omega_d < 5.5201$. To highlight the propagation of the revivals, we mark the wave crests with orange and red symbols, which track the evolution both in time and across this fictitious parameter space. This representation clearly illustrates how the revivals emerge in one region of parameter space (small $h$) and then spread, providing a visual and quantitative understanding of the dynamics underlying the stroboscopic memory and revival behavior in the driven PXP system. The presence of this highly controllable mechanism is intriguing, as it allows to tune the parameters to determine, on demand, the specific times at which revivals occur. This behavior can be understood in terms of the deformation of the Floquet spectrum induced by tuning the driving parameters, in particular through changes in its overlap with the Nèel state.

\section{Spectral origin of the mechanism}
\label{sec:spectral_origin}
We now discuss the properties of the Floquet spectrum, focusing on the overlaps with the Nèel state, which play a central role in the dynamical mechanism described above.

\begin{figure}[tp]
\centering
\includegraphics[width=0.92\linewidth]{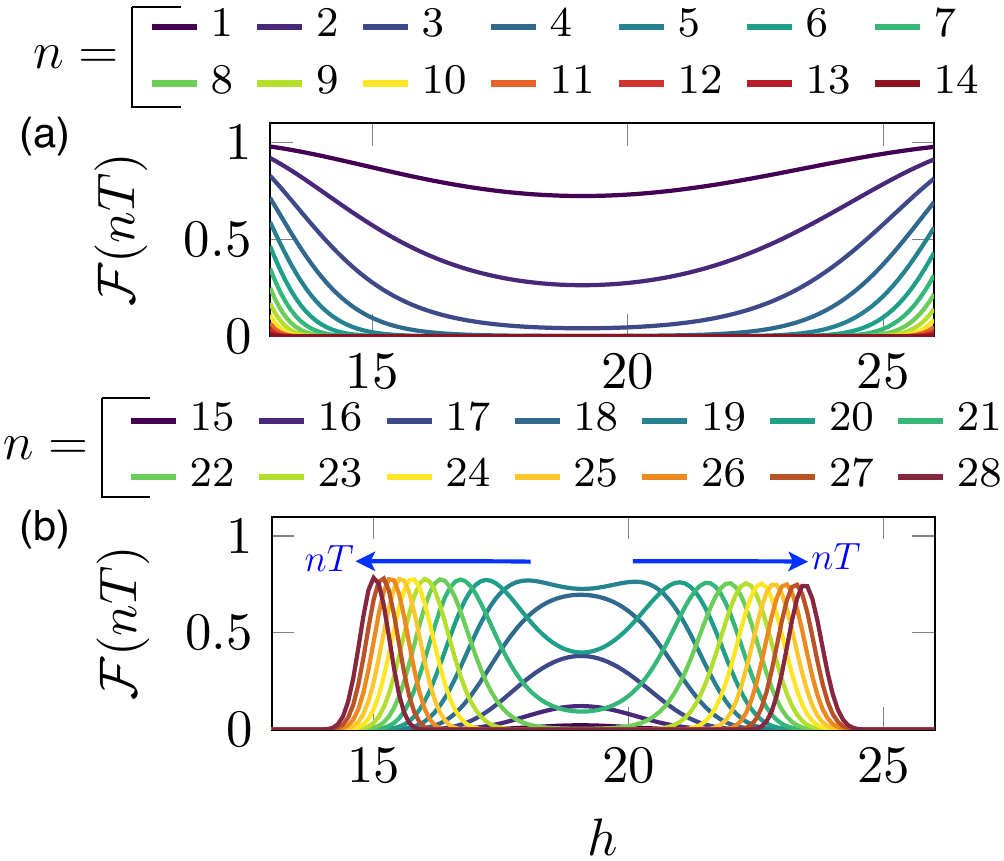}
\caption{Revival mechanism in the region of $h/\omega_d$ in between the two zeros of $J_0(h/\omega_d)$: $2.4048<h/\omega_d < 5.5201$. We consider $\omega_d=5$ and $L=12$. Panel (a) shows the decay phase, where the fidelity monotonically decreases. Panel (b) shows the revival phase, where the fidelity grows in time following, in $h$, the trajectories indicated by the blue arrows. The blue arrow guides the eye along the time evolution of the revivals during the revival phase.}
\label{fig:central}
\end{figure}
In Figs.~\ref{fig:spectral_fig}(a)–(d) we show the overlaps of the Nèel state with the Floquet spectrum across four distinct regions of the driving amplitude at fixed $\omega_d=5$. The four regimes considered correspond to the following ranges of driving amplitudes: small ($h \in [2.45, 2.85]$), moderate ($h \in [5.05, 5.45]$), high ($h \in [7.45, 7.85]$), and very high ($h \in [9.6, 10]$), close to the FSN.  Despite their differences, these regimes exhibit a common structure: an arc of dominant states, reminiscent of the QMBS present in the undriven model, together with the remaining states forming the bulk. This separation becomes increasingly pronounced as $h$ is increased toward the FSN. We note that the states belonging to the arc exhibit a spacing that depends on the driving amplitude. On the one hand, increasing the driving amplitude reduces the Floquet bandwidth and therefore decreases overall the spacing between the Floquet eigenstates. On the other hand, we observe that states near the center of the Floquet spectrum are less sensitive to $h$, whereas those near the edges are more sensitive to it. Thus, $h$ also controls the spacing in specific regions of the Floquet spectrum. In general, as $h$ is increased toward the FSN, this sensitivity becomes more pronounced. In all cases, we identify the Floquet energy separation $\Delta \epsilon^F$ between the zero–quasi-energy state and its nearest neighbor on the arc as the dominant energy scale, we pictorially represent it in Figs.~\ref{fig:spectral_fig}(a)–(d). Far from the FSN, this quantity approximates the characteristic spacing of all states belonging to the arc. Closer to the FSN, the states at the edges of the arc develop reduced overlap, and $\Delta \epsilon^F$ remains the relevant energy scale separating the dominant states. 

The latter energy scale is expected to determine the time at which revivals occur, which we define as $T_{\rm rev} \coloneqq 2\pi/\Delta\epsilon^F$. Since we focus on the stroboscopic dynamics, we introduce
\begin{equation}
n_{\rm rev}= \frac{T_{\rm rev}}{T} \coloneqq \frac{\omega_d}{\Delta\epsilon^F},
\end{equation}
which specifies the multiples of the driving period at which revivals are expected to appear. In Figs.~\ref{fig:spectral_fig}(e) and (f), we report the behavior of $n_{\rm rev}$ as a function of $h/\omega_d$ for different system sizes $L$, with $\omega_d=5$ (panel (e)), and for different values of $\omega_d$, with $L=10$ (panel (f)). The behavior of $n_{\rm rev}$ is well fitted by
$n_{\rm rev} = \frac{\omega_d}{\gamma J_0(h/\omega_d)} + \alpha$,
where the fit parameters $\gamma$ and $\alpha$ depend on $\omega_d$ (see Appendix~\ref{app:fit}). For $h/\omega_d$ approaching the first zero of the Bessel function, we have
$n_{\rm rev} \approx \frac{\omega_d}{\gamma J_0(h/\omega_d)}$,
indicating that $\Delta\epsilon^F \propto J_0(h/\omega_d)$. This shows that the compression of the dominant level spacing follows the compression of the entire Floquet bandwidth, providing a useful tool for understanding the revival mechanism. 

Having the Floquet spectral structure of the system in mind, we now discuss the evolution of $\mathcal{F}(nT)$ at fixed $\omega_d$, as shown in Figs.~\ref{fig:spectral_fig}(h),(i). We consider $h<2.4048 \omega_d$. We identify two distinct phases: the decay phase (panel (h)), where the fidelity steadily decreases, and the revival phase (panel (i)), where the fidelity exhibits characteristic revival peaks. In the decay phase, $\mathcal{F}(nT)$ gradually decreases, forming a bell-like structure around the FSN, where the dynamics is particularly slow. In the revival phase, the fidelity shows sharp peaks corresponding to the revival of the system’s state. The revivals follow the mechanism introduced in Fig.~\ref{fig:memory1} and can be better appreciated by considering consecutive stroboscopic times. The first significant revival in the considered $h$ interval appears at $n=8$ for small $h$, with a distribution in $h$ of considerable width. Over time, the center of this distribution shifts right toward the FSN, while its width shrinks, as if the FSN acts as a barrier that drags the evolution. This apparent motion arises because $\Delta\epsilon^F$ decreases with increasing $h$, while the narrowing of the distribution is due to the varying sensitivity of the Floquet spectrum to $h$ in different regions: when the spectrum is weakly sensitive (small $h$), revivals occur at specific times over a wider range of $h$, whereas when the spectrum is highly sensitive (near the FSN), the range is narrower. This feature is important for controlling dynamics: a larger width allows more tolerance in $h$ calibration for a desired response time, while a smaller width requires higher precision. Finally, we note that, at $t = 15T$, the fidelity distribution
develops an emergent tail at small h, signaling the regular
appearance of the second revivals.

Subsequently, we analyze the revival mechanism in the region $2.4048 \omega_d < h < 5.5201 \omega_d$, at fixed $\omega_d$, as shown in Fig.~\ref{fig:central}(a),(b). In this region, the dynamics follow a markedly different pattern. First, the decay phase (panel (a)) persists for longer times: no significant revival occurs before a time that is roughly twice the minimal revival time observed in the previous region, indicating a slower initial relaxation of the fidelity.
The revival phase (panel (b)) exhibits a particularly interesting mechanism. Initially, a “blob” forms at the center of the $h$ interval, corresponding to the range of $h$ values where the system first shows signs of revival. As time progresses, the height of this blob grows, reaching a maximum when the revival is strongest. After reaching this maximum, the blob splits into two separate parts that move in opposite directions in the fictitious $h$ space, each approaching one of the two FSN regions. During this process, the width of each part gradually shrinks. The splitting and motion of the blob reflect that different regions of the Floquet spectrum respond symmetrically with respect to the center of the region, resulting in revivals that propagate toward the two FSN in opposite directions. 
We further note that, for times up to $t \leq 28T$, no second revivals are observed, confirming that the dynamics in this region are significantly slower than in the previous case. This slowdown is a direct consequence of the Floquet spectrum structure, in which the bandwidth decreases in magnitude across different regions separated by the FSN (see Fig.~\ref{fig:memory1}(a)).
\section{Revival mechanism interpolating between polarized and Nèel states}
\label{sec:theta}
In this section, we investigate how the revival mechanism depends on the choice of initial state. Specifically, we track the evolution of the mechanism by interpolating between the polarized and Nèel states using the $\ket{\Theta_+}$ state:
\begin{align}
\ket{\Theta_+} = \ket{\theta_+} \otimes \ket{0} \otimes \ket{\theta_+} \otimes \ket{0} \otimes \ldots,
\end{align}
where $\ket{\theta_+} \coloneqq \cos(\theta)\ket{0} + \sin(\theta)\ket{1}$. By varying $\theta$ in the interval $[0,\pi/2]$, the state smoothly interpolates between the polarized state, $\ket{\boldsymbol{0}} = \ket{0000\ldots}$, and the Nèel state, $\ket{\mathbb{Z}_2} = \ket{1010\ldots}$. We perform our analysis by fixing $\omega_d$.

\begin{figure}[t!]
\centering
\includegraphics[width=\linewidth]{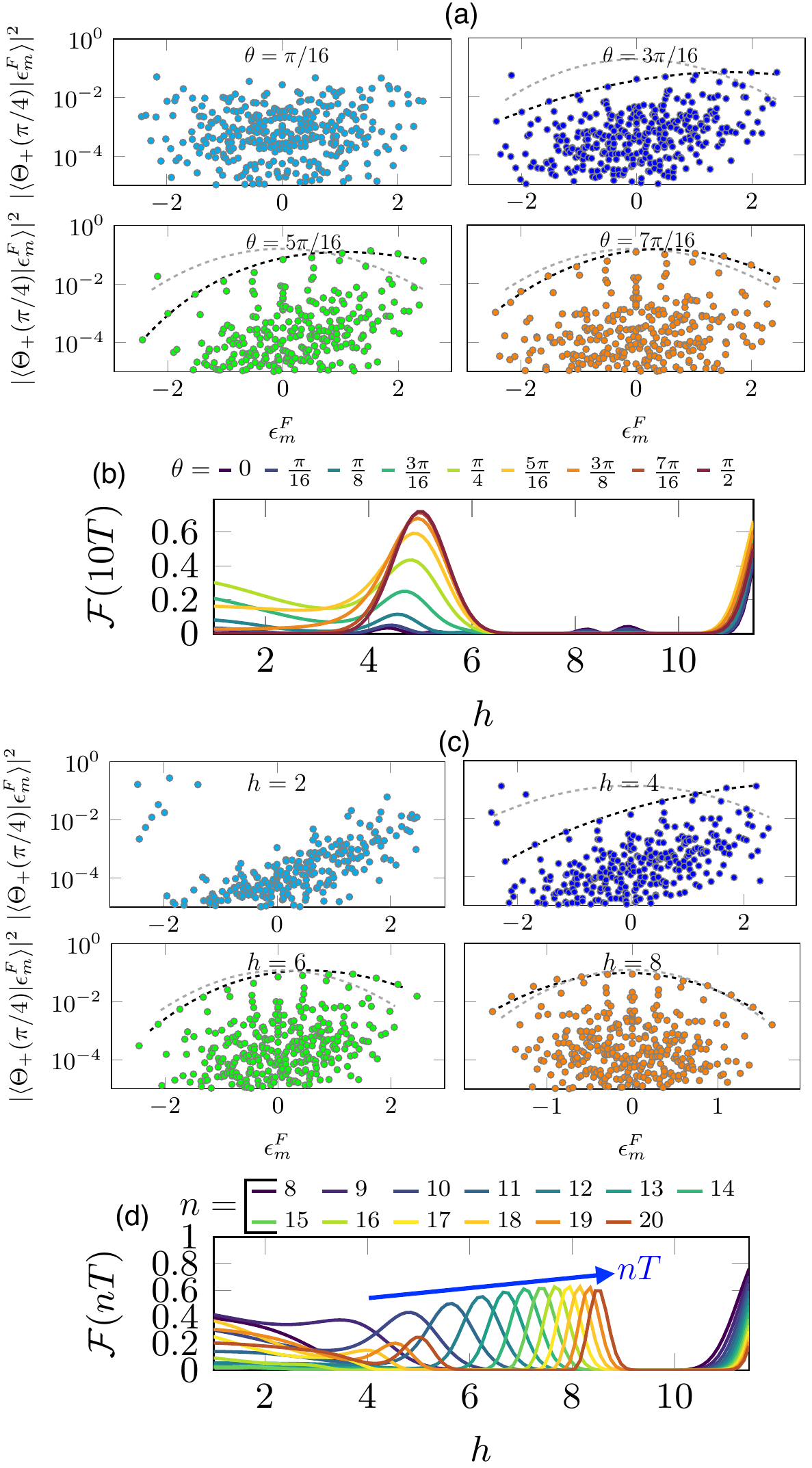}
\caption{Revival mechanism for the $\ket{\Theta_+}$ state with $\omega_d=5$ and $L=12$. Panel (a): overlap between the $\ket{\Theta_+}$ state and the Floquet eigenstates for different values of $\theta$ (different colors), computed at fixed $h = 5$. Panel (b): fidelity between the initial state $\ket{\Theta_+}$ and its stroboscopic time-evolved state at $t = 10T$, shown as a function of $h$ for several values of $\theta$ in the range $0$ to $\pi/2$. Panel (c): overlap between the $\ket{\Theta_+}$ state and the Floquet eigenstates for different values of $h$ (different colors), computed at fixed $\theta = \pi/4$. Panel (d): fidelity for $\theta = \pi/4$ evaluated at different stroboscopic times $nT$, with $n$ indicated in the legend. The blue arrow guides the eye along the time evolution of the revivals during the revival phase. In panels (a) and (c) the dashed black lines are guides to the eye highlighting the emergence of arc-like structures, while the dashed gray line indicates the corresponding arc structure for the Néel-state case.}
\label{fig:theta}
\end{figure}
\begin{figure*}[!t]
\centering
\includegraphics[width=\linewidth]{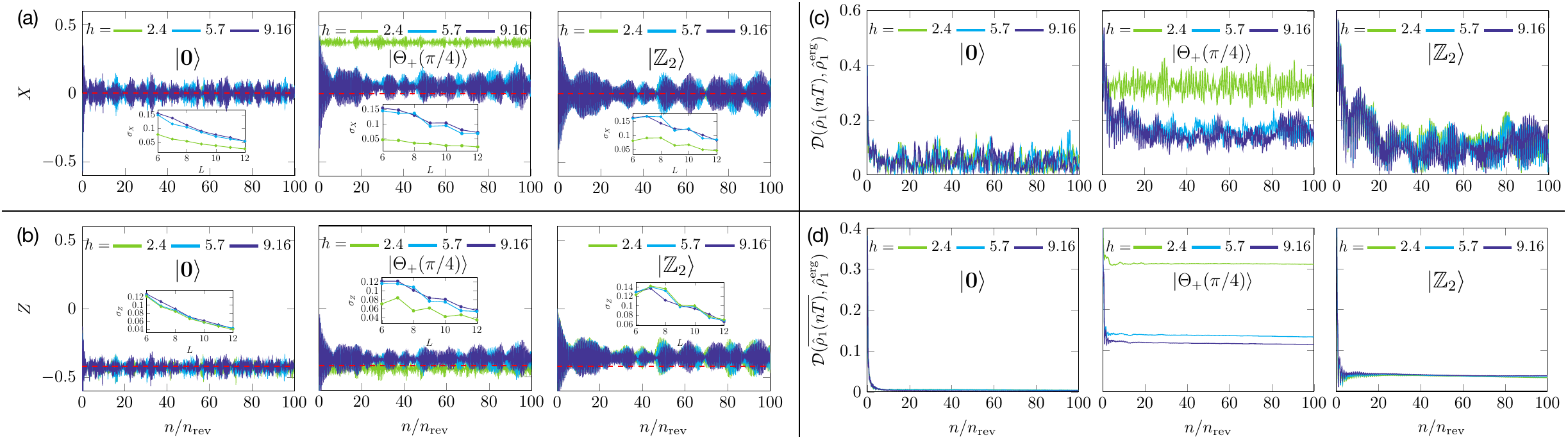}
\caption{Analysis of Floquet thermalization for different initial states ($\ket{\boldsymbol{0}},\,\ket{\Theta_+(\pi/4)},\,\ket{\mathbb{Z}_2}$) and different driving amplitude values, the latter are $h=2.4, 5.7, 9.14$ and have revival times that are approximately integer multiples of the period: $n_{\rm rev}=\omega_d/\Delta\epsilon^F\approx 8, 11,25$. The driving frequency is fixed to $\omega_d=5$ and $L=12$. Panels (a) and (b): dynamics at stroboscopic times of $X(t)$ (a) and $Z(t)$ (b) for the three initial states and the three values of the driving amplitude, compared with their ergodic expectations $X^{\rm erg}=0$ and $Z^{\rm erg}$ in dashed red. The insets report the standard deviations of the signals, defined as $\sigma_A=\sqrt{\frac{1}{100 n_{\rm rev}}\sum_{n=1}^{100 n_{\rm rev}}\left(A(nT)-\frac{1}{100 n_{\rm rev}}\sum_{m=1}^{100 n_{\rm rev}} A(mT)
\right)^2}$, with $A=X,Z$. Panel (c) reports the trace distance between first-site density matrix at stroboscopic times $\hat{\rho_1}(nT)$ and the corresponding ergodic density matrix $\hat{\rho}_1^{\rm erg}$, while panel (d) reports the distance between the latter and the time-averaged first-site state $\overline{\hat{\rho_1}(nT)}$.}
\label{fig:thermalization}
\end{figure*}
We begin by analyzing the overlap of the initial state with the Floquet spectrum. Fixing $h$, we study how the overlap structure evolves for different values of $\theta \in [0, \pi/2]$, see Fig.~\ref{fig:theta}(a). For small $\theta$, the overlap predominantly reflects a bulk-like configuration in line with that of a polarized state. As $\theta$ increases, the structure gradually evolves, exhibiting the characteristic arc of nearly equally spaced states seen above the bulk in the Néel state. This similarity can be visually appreciated by comparing the emergent arc in these plots, highlighted with a black dashed line, to the reference arc of the Néel state, shown in gray, with the black arc gradually converging toward the gray one as $\theta$ approaches $\pi/2$.

To understand how this spectral deformation influences dynamics, we examine the stroboscopic fidelity at fixed time $t=10T$ for $h < 2.4048 \, \omega_d$, see Fig.~\ref{fig:theta}(b). In the small–$\theta$ regime, the fidelity remains low and displays only shallow humps. These features result from residual finite-size effects, which also prevent full ergodicity yet maintaining broadly ergodic-like dynamics; the ergodic properties of the system are analyzed in more detail in section~\ref{sec:thermalization}. Increasing $\theta$ qualitatively changes the behavior: a pronounced central peak appears, signaling the onset of revivals as the Floquet spectrum develops the arc structure with nearly equally spaced states. The height of this peak grows steadily with $\theta$ and reaches its maximum at $\theta = \pi/2$ (Nèel state), indicating that this configuration provides the clearest and most robust revival signal. Additionally, at intermediate angles, the fidelity distributions develop long tails at small $h$, indicating the presence of memory of the initial state with moderate strength across a wide range of driving amplitudes. In contrast, the limiting cases ($\theta = 0$ and $\theta = \pi/2$) produce either absent or sharply localized fidelity profiles, showing rapid loss of memory or confinement to narrow parameter regions. Thus, intermediate $\theta$ offers a tunable handle for controlling the persistence of initial state information.

Next, we focus on the intermediate case $\theta = \pi/4$ to investigate the effect of varying $h$ on both the Floquet spectrum and dynamics. The overlap of $\ket{\Theta_+(\pi/4)}$ with Floquet eigenstates is shown in Fig.~\ref{fig:theta}(c). Unlike the Nèel state, whose overlap structure remains largely insensitive to $h$ (see Fig.~\ref{fig:spectral_fig}(a)-(d)), the intermediate-$\theta$ state exhibits a strong $h$ dependence, indicating access to qualitatively distinct spectral and dynamical regimes. At small $h$, the overlap displays an approximately exponentially growing pattern, with a set of isolated low-quasi-energy states occupying the top-left region of the plot. Remarkably, as $h$ approaches the FSN, the overlap recovers a Nèel-like configuration, demonstrating that, at intermediate $\theta$, tuning $h$ allows exploration of two fundamentally different spectral structures. This is evident from the emergent correspondence between the arc patterns in these plots (black dashed line) and those of the Nèel state (gray dashed line) which are reported as reference. We note that, whereas in the case shown in Fig.~\ref{fig:theta}(a) the correspondence emerges from the fact that $\theta$ interpolates between the polarized and Nèel states, here the result is less intuitive and more striking. It demonstrates that the driving parameters can be appropriately tuned to generate Nèel-like dynamics even from different initial states.

To illustrate this, we fix $\theta = \pi/4$ and track the evolution of the revival peaks as a function of $h$, as shown in Fig.~\ref{fig:theta}(c). Apparently, the overall mechanism resembles that of the Nèel state: peaks first emerge at small $h$ and propagate toward the FSN, effectively being “dragged” along the $h$ axis while gradually narrowing. However, a key distinction arises. At small $h$, the dynamics is irregular and fluctuating, due to an overlap structure that differ with the one of the Nèel state. This irregularity gradually fades as the peaks approach the FSN, where the dynamics becomes increasingly regular and Nèel-like. Consequently, at intermediate $\theta$, the system exhibits a coexistence of two dynamical “characters”: erratic, fluctuating behavior at small $h$, and well-structured revivals at larger $h$. These two distinct regimes produce relevant differences even at long times and correspond to separate mechanisms through which the system avoids thermalization, as discussed in detail in section~\ref{sec:thermalization}. This unified perspective directly links the spectral deformation induced by varying $h$ to the observed dynamical patterns, highlighting the rich interplay between initial state choice, spectral structure, and dynamical response.
\section{Non-thermal Floquet dynamics and initial state dependence}
\label{sec:thermalization}
In the previous sections, we have analyzed the revival dynamics of the driven PXP model initialized in the Nèel state, as well as for initial states interpolating between it and the fully polarized state. A natural question concerns the ergodic properties of this dynamics. To address it, we look at the long-time dynamics of the system and we study thermalization at the level of a single site, which we treat as a subsystem, while the rest of the chain plays the role of an effective bath that may induce thermalization. Periodically driven quantum systems are expected to thermalize when the expectation values of local observables agree with those computed in the infinite-temperature state
$\hat{\rho}^{\rm erg}=\hat{\mathds{1}}_{D}/D$,
where $D$ denotes the Hilbert-space dimension~\cite{seetharam2018absence, staszewski2025krylov, lazarides2014equilibrium, hou2025floquet, dalessio2014long, mori2018thermalization}. In our case, the relevant ergodic state is instead identified as
$\hat{\rho}^{\rm erg}=\hat{\mathds{1}}_{F_{L+2}}/F_{L+2}$,
since the dynamics is restricted to the blockaded subspace of dimension $F_{L+2}$.

To characterize the ergodic features of the dynamics, we quantify thermalization using the trace distance between the reduced density matrix of the subsystem and the corresponding ergodic state. In particular, we consider the following quantities:
\begin{align}
&\mathcal{D}(\hat{\rho}_S(t),\hat{\rho}_S^{\rm erg})=\frac{1}{2}\|\hat{\rho}_S(t) - \hat{\rho}_S^{\rm erg} \|_1, \label{eq:inst_dist}
\\
&\mathcal{D}\left(\overline{\hat{\rho}_S(t)},\hat{\rho}_S^{\rm erg}\right)=\frac{1}{2}\|\overline{\hat{\rho}_S(t)} - \hat{\rho}_S^{\rm erg} \|_1;\label{eq:avg_dist}
\end{align}
Here, $\hat{\rho}_S = \operatorname{Tr}_{\bar{S}}[\hat{\rho}]$, with $\bar{S}$ denoting the complement of $S$, and $\overline{\hat{\rho}_S(t)} = \frac{1}{t} \int_0^t dt' \hat{\rho}_S(t')$. These two quantities distinguish between strong and weak thermalization~\cite{banulus2011strong}. Strong thermalization occurs when both distances vanish at long times. If only the time-averaged distance vanishes while the instantaneous distance continues to oscillate, the system exhibits weak thermalization.

Since we focus on the case where $S$ is a single site, the trace distances reduce to distances between Bloch vectors associated with the corresponding single-site density matrices:
\begin{align}
        &\mathcal{D}(\hat{\rho}_j(t),\hat{\rho}_j^{\rm erg})=\frac{1}{2} |\boldsymbol{r}_j(t) - \boldsymbol{r}_j^{\rm erg}|;\label{eq:dist_1site}
        \\& \mathcal{D}(\overline{\hat{\rho}_j(t)},\hat{\rho}_j^{\rm erg})= \frac{1}{2} \left|\overline{\boldsymbol{r}_j(t)} - \boldsymbol{r}_j^{\rm erg}\right|;\label{eq:timeavgdist_1site}
\end{align}
The Bloch vectors are given by $\boldsymbol{r}_j(t)=\left(\braket{\hat{X}_j(t)},\braket{\hat{Y}_j(t)},\braket{\hat{Z}_j(t)}\right)$, $\overline{\boldsymbol{r}_j(t)}=\left(\overline{\braket{\hat{X}_j(t)}},\overline{\braket{\hat{Y}_j(t)}},\overline{\braket{\hat{Z}_j(t)}}\right)$, and $\boldsymbol{r}_j^{\rm erg}=(0,0,Z_j^{\rm erg})$. The nonzero $z$-component of the ergodic Bloch vector reflects the constraint imposed by the blockaded Hilbert space and is given by $Z_j^{\rm erg}=\frac{F_{j}F_{L-j+1}- F_{j+1}F_{L-j+2}}{F_{L+2}}$ (see appendix~\ref{app:zjerg}). Thus, the single-site thermalization passes through the convergence of local observables $\braket{\hat{X}_j(t)},\braket{\hat{Y}_j(t)},\braket{\hat{Z}_j(t)}$ and their time-averages to respectively $0,\,0,\,Z_j^{\rm erg}$.

We begin by examining the thermal behavior of the system at the level of the entire chain. To this end, we define the spatially averaged observables
$X(t)=\frac{1}{L}\sum_{j=1}^L \braket{\hat{X}_j(t)}$,
$Y(t)=\frac{1}{L}\sum_{j=1}^L \braket{\hat{Y}_j(t)}$, and
$Z(t)=\frac{1}{L}\sum_{j=1}^L \braket{\hat{Z}_j(t)}$. In Fig.~\ref{fig:thermalization}(a),(b) we show the time evolution of $X$ and $Z$, while the dynamics of $Y$ is reported in appendix~\ref{app:Ydynamics}, Fig.~\ref{figapp:longtimeY}. We fix $\omega_d=5$, consider the regime $h<2.4048\omega_d$ and focus on three representative values of $h$. These values are chosen such that $n_{\rm rev}=\omega_d/\Delta\epsilon^F$ is approximately integer, and correspond, respectively, to parameters far from the FSN ($h=2.4$), at an intermediate location ($h=5.7$), and close to ($h=9.14$) the FSN.

We first consider the dynamics of $X$ (Fig.~\ref{fig:thermalization}(a)) and observe clear differences between the three initial states. For the polarized and Néel states, the dynamics is largely insensitive to $h$: the polarized state exhibits noisy fluctuations around the ergodic value $X^{\rm erg}=0$, while the Néel state shows more coherent oscillations, manifested in the revivals previously discussed. In contrast, the interpolating state $\ket{\Theta_+(\pi/4)}$ displays a strong $h$-dependence. For the smallest $h$, $X(t)$ weakly fluctuates and remains far from the ergodic value; for the other two $h$ values, the dynamics approaches but does not fully reach the ergodic expectation. This behavior reflects the $h$-dependent overlap of the initial state with the Floquet spectrum and indicates that, on average, the interpolating state fails to thermalize. The $Y$ dynamics (Fig.~\ref{figapp:longtimeY}) shows comparable trends, with the difference that $h=2.4$ and $h=5.7$ produce similar dynamics, while $h=9.14$ does not. For the polarized and Néel states, $Y(t)$ oscillates either noisily or more coherently around $Y^{\rm erg}=0$ for all $h$. Whether these states are ergodic or non-ergodic is revealed by the dynamics of $Z$. In Fig.~\ref{fig:thermalization}(b) we show the dynamics of $Z$ for the three initial states. The interpolating state is weakly dependent on $h$ and consistently deviates from the ergodic value. For the polarized and Néel states, the distinction between noisy and coherent behavior persists; additionally, the polarized state oscillates around the ergodic expectation $Z^{\rm erg}=\frac{1}{L}\sum_j Z_j^{\rm erg}$, while the Néel state clearly deviates from it, even on average, indicating non-ergodic behavior. 

The insets of Figs.~(a) and (b) report the fluctuations, quantified by the standard deviation, of the signals $X(nT)$ and $Z(nT)$ over the time interval $n\in[0,100 n_{\rm rev}]$ for different values of the system size $L$. While the polarized state exhibits a regular decay with increasing $L$, suggesting the onset of strong thermalization in the thermodynamic limit, the Néel state and the interpolating $\ket{\Theta_+(\pi/4)}$ state display irregular behaviors as a function of $L$. This is indicative of persistent oscillations in the system, which prevent the identification of a clear scaling in $L$ within a finite time window. We also note that, overall, fluctuations are reduced far from the FSN; in particular, in many cases the curve corresponding to $h=2.4$ is significantly smaller than the others.

Finally, we analyze the instantaneous and time-averaged distances defined in Eqs.~\eqref{eq:dist_1site} and~\eqref{eq:timeavgdist_1site}. For the instantaneous distance, the polarized state is the closest to the ergodic value, with small fluctuations due to the finite system size. Larger instantaneous distances are observed for the Néel and interpolating states. In particular, the interpolating state stays far from zero throughout the dynamics, never locally resembling the ergodic state. For $h$ far from the FSN, this distance is especially large, reflecting a regime far from thermalization, due to the significant contributions of $X(t)$ and $Y(t)$. The time-averaged distance confirms these trends. The Néel and interpolating states avoid ergodicity in distinct ways, while the polarized state approaches the ergodic expectation on average. Although the finite system size prevents a sharp distinction between weak and strong thermalization, we can assert that the polarized state exhibits at least weak thermalization, whereas the other two states do not.

To conclude, two different forms of non-ergodicity emerge. For the Néel state, the long-time-averaged local state has, on average, zero $X$ and $Y$ expectation values, but a non-zero $Z$ expectation value, causing it to deviate from the infinite-temperature prediction. For the interpolating state, particularly far from the FSN, the long-time-averaged state exhibits non-zero $X$ and $Y$ components, further enhancing the athermality of the system. This behavior is associated with a complex overlap structure with the Floquet states (Fig.~\ref{fig:theta}(c)). As the system approaches the FSN, the interpolating state's overlap structure evolves toward an arc-like pattern, and its long-time-averaged state becomes similar to that of the Néel state, with $X$ and $Y$ expectation values approaching zero on average, yet still remaining non-ergodic.

\section{Conclusions $\&$ outlook}
\label{sec:conclusions}
We investigated the mechanism underlying the emergence of revivals in the non-ergodic dynamics of the periodically driven PXP model. Our analysis employed Floquet theory, focusing on the stroboscopic evolution generated by the Floquet operator.

The structure of the Floquet quasi-energy spectrum exhibits special regions in parameter space where the spectrum significantly narrows, which we refer to as Floquet spectrum narrowing (FSN) regions. In these regions, the stroboscopic dynamics slows down abruptly. FSN regions play a central role in shaping the emergence, evolution, and robustness of revivals when the system is initialized in the Néel state.

The dynamics can be interpreted as an effective wave packet propagating in the driving-parameter space, with FSN regions acting as barriers that delay its motion. This leads to a harmonic evolution governed by the Floquet spectral structure and its overlap with the Néel state. Notably, the arc-like structure of this overlap, reminiscent of static quantum many-body scars (QMBS), is preserved under periodic driving. As a result, a dominant quasi-energy spacing emerges, which sets the timescale and regularity of the observed revivals.

We subsequently extended the analysis to initial states that interpolate between the Néel state and the fully polarized state, revealing a different mode of response to periodic driving. These states exhibit a hybrid behavior that can be controlled through the driving parameters. For instance, we showed that, by tuning the driving amplitude, one can interpolate between two distinct revival mechanisms. In this case, the structure of the overlap between the initial state and the Floquet spectrum is highly sensitive to the drive, in contrast to the Néel state, whose overlap structure remains more robust.

Finally, we analyzed the long-time dynamics to investigate whether and how the system conforms to (or avoids) Floquet thermalization, which predicts convergence to infinite-temperature states. For the system sizes considered, we find that the fully polarized state weakly thermalizes, meaning that it converges on average to the ergodic state. In contrast, the Néel state avoids thermalization, reaching at long times a time-averaged state that exhibits zero $X$ and $Y$ expectation values on average, but non-ergodic $Z$ expectation values, making it inconsistent with infinite-temperature predictions. Interestingly, the hybrid character of the interpolating states is also manifested in their long-time dynamics, which can follow diverse pathways to evade thermalization. In particular, we find that in certain regimes, the long-time-evolved state can develop pronounced non-zero $X$ and $Y$ expectation values, further enhancing its non-thermal character and representing a second mechanism by which the system evades thermalization.

This work highlights the presence of a controlled mechanism that governs the emergence of revivals and prevents thermalization in the driven PXP model. The ability to tune dynamical features, including the timescales associated with different initial states, opens promising avenues for applications such as quantum information storage and the controlled manipulation of coherent many-body dynamics on demand. Among the potential directions arising from this work, one could explore the mechanism under different types of driving, including quasi-periodic protocols~\cite{dutta2025prethermalization}. Furthermore, the robustness and possible deformation of various non-ergodic dynamical regimes, such as local reminiscent dynamics~\cite{perciavalle2025local}, can be systematically investigated and tested under periodic driving.

\section{Acknowledgments}
This work was partially funded by the
PNRR MUR Project No. PE0000023-NQSTI through the
secondary projects QuCADD, ThAnQ and QuSOE. Numerical simulations are performed by using the Julia package \texttt{QuantumOptics.jl}~\cite{kramer2018quantum}.

\appendix

\section{Analysis of the expected revival time}
\label{app:fit}
In this section, we analyze in detail the behavior of $n_{\rm rev}$, paying particular attention to its dependence on $h$ for different values of $\omega_d$. We focus on the region of driving parameters such that $h/\omega_d < 2.4048$. Assuming that the Floquet bandwidth follows $J_0(h/\omega_d)$ and that the dominant quasi-energy spacing $\Delta\epsilon^F$ deforms proportionally to the bandwidth, we may posit that $\Delta\epsilon^F = \delta\, J_0(h/\omega_d)$, where $\delta$ is a proportionality factor, implying a direct proportionality between $\Delta\epsilon^F$ and the Bessel function. This, in turn, would naturally lead to $n_{\rm rev} = \omega_d / (\delta \, J_0(h/\omega_d))$. However, as we will show, this assumption is not completely accurate. In fact, fixing $\omega_d$ and inspecting the Floquet spectrum, we observe that, as shown in Fig.~\ref{fig:memory1}(a), for small $h/\omega_d$ the spectrum is not shaped by the Bessel function $J_0(h/\omega_d)$, making the simple proportionality ansatz inaccurate. For this reason, we assume that the behavior is corrected by an offset and write
\begin{equation}
n_{\rm rev}(h)=\frac{\omega_d}{\gamma J_0(h/\omega_d)} + \alpha.
\label{eq:fit2}
\end{equation}
For $h\ll\omega_d$, we have $J_0(h/\omega_d)\approx 1$, so $n_{\rm rev} \approx \frac{\omega_d}{\gamma} + \alpha$, while when approaching the FSN we have $\frac{\omega_d}{\gamma J_0(h/\omega_d)} \gg \alpha$, leading to $n_{\rm rev} \approx \frac{\omega_d}{\gamma J_0(h/\omega_d)}$. In the latter regime, the hypothetical validity of the fit confirms the correspondence between the deformation of the Floquet spectrum and the deformation of the dominant quasi-energy spacing.

\begin{figure}[!t]
\centering
\includegraphics[width=\linewidth]{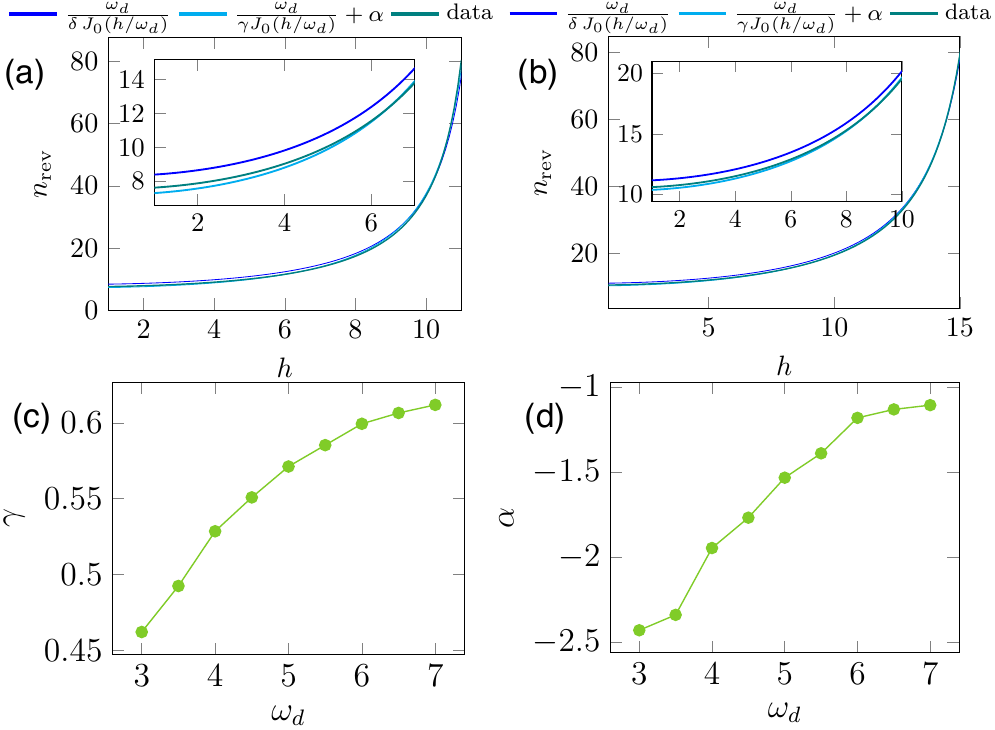}
\caption{Panels (a) and (b): comparison between $n_{\rm rev}$ extracted from the Floquet spectrum (data) and two different fits: $\frac{\omega_d}{\gamma J_0(h/\omega_d)} + \alpha$ and $\frac{\omega_d}{\delta, J_0(h/\omega_d)}$. Panel (a) shows the results for $\omega_d=5$, while panel (b) shows the results for $\omega_d=7$. The system size is fixed to $L=10$ in both panels. Panels (c) and (d): fit parameters $\gamma$ and $\alpha$ as a function of $\omega_d$, obtained by fitting $n_{\rm rev} = \frac{\omega_d}{\gamma J_0(h/\omega_d)} + \alpha$. The size of the system is $L=10$.}
\label{fig:fit_figure}
\end{figure}

\begin{table}[ht!]
\centering
\begin{tabular}{|c|c|c|}
  \hline
  $\omega_d$ & $\gamma$ & $\alpha$ \\
  \hline
  3   & $0.46215  \pm 0.0116$   & $-2.42885 \pm 0.4577$ \\
  3.5 & $0.492468 \pm 0.00334$  & $-2.33842 \pm 0.1314$ \\
  4   & $0.528536 \pm 0.00267$  & $-1.94613 \pm 0.1045$ \\
  4.5 & $0.550855 \pm 0.002316$ & $-1.76802 \pm 0.09638$ \\
  5   & $0.57121  \pm 0.001885$ & $-1.53336 \pm 0.08017$ \\
  5.5 & $0.585289 \pm 0.001609$ & $-1.39035 \pm 0.07179$ \\
  6   & $0.599358 \pm 0.00126$  & $-1.18218 \pm 0.05604$ \\
  6.5 & $0.606491 \pm 0.00115$  & $-1.13246 \pm 0.05498$ \\
  7   & $0.611822 \pm 0.001078$ & $-1.10751 \pm 0.05569$ \\
  \hline
\end{tabular}
\caption{Fit parameters $\gamma$ and $\alpha$ for different driving frequencies $\omega_d$, with their respective uncertainties.}
\label{tab:fit_parameters}
\end{table}
In Fig.~\ref{fig:fit_figure}(a),(b) we compare the behavior of $n_{\rm rev}(h)$ for two fixed values of $\omega_d$ with the two proposed fits. In both cases, we observe that the fit given by Eq.~\eqref{eq:fit2} works reasonably well, especially at small $h$. Consequently, we consider fixed values of $\omega_d \in [3,3.5,4,\ldots,7]$ and, for each $\omega_d$, we fit $n_{\rm rev}(h)$ data in the interval $h \in [1,2.2048\omega_d]$, i.e., before the first zero of the Bessel function. The two fit parameters are plotted in Fig.~\ref{fig:fit_figure}(c),(d) as a function of $\omega_d$ and are also reported with their respective errors in Table~\ref{tab:fit_parameters}. Overall, the fit works better for larger $\omega_d$, with the errors on the parameters decreasing accordingly. At the same time, the two fit parameters tend to stabilize for large $\omega_d$. Finally, we note that this approach allows us to approximately identify the minimal time required for observing a revival at each $\omega_d$. Indeed, since $n_{\rm rev}$ is a positive monotonically increasing function of $h$ in the region of parameters considered, the minimal revival time is obtained for $J_0(h/\omega_d)\approx 1$, giving $n_{\rm rev}^{\rm min} \approx \frac{\omega_d}{\gamma} + \alpha$. For instance, for $\omega_d=7$, we have $n_{\rm rev}^{\rm min} \approx 10.3325$.
\section{Evaluation of $Z_j^{\rm erg}$}
\label{app:zjerg}
Thermalizing driven closed quantum systems are characterized by long-time expectation values of local operators that coincide with those of the infinite-temperature mixed state. In the constrained Hilbert space of dimension $D_L = F_{L+2}$, this state reads 
\begin{equation}
\hat{\rho}_{[1,\ldots,L]}^{\rm erg}=\frac{\hat{\mathds{1}}_{F_{L+2}}}{F_{L+2}}=    \frac{1}{D_L}\sum_{\boldsymbol{s}\in \mathcal{B}_L}\ket{\boldsymbol{s}}_{[1,\ldots,L]}\bra{\boldsymbol{s}};    
\end{equation}
where $\mathcal{B}_L$ denotes the constrained Hilbert space and the subscript $[1,\ldots,L]$ indicates that the state is defined over the entire chain. In general, the subscript $[\ell_1,\ldots,\ell_2]$ specifies that the state has support on the region between sites $\ell_1$ and $\ell_2$ of the chain.

\begin{figure*}[!t]
\centering
\includegraphics[width=0.65\linewidth]{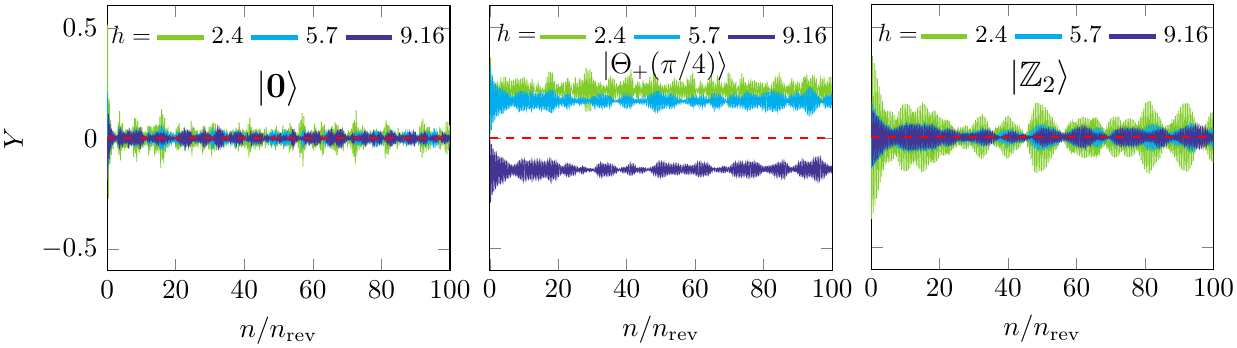}
\caption{Analysis of the long time dynamics of $Y(t)$ for different initial states ($\ket{\boldsymbol{0}},\,\ket{\Theta_+(\pi/4)},\,\ket{\mathbb{Z}_2}$) and different driving amplitude values; the latter are $h=2.4, 5.7, 9.14$ and have revival times that are approximately integer multiples of the period: $n_{\rm rev}=\omega_d/\Delta\epsilon^F\approx 8, 11,25$. The driving frequency is fixed to $\omega_d=5$ and $L=12$.}
\label{figapp:longtimeY}
\end{figure*}
This ergodic state can be decomposed recursively as
\begin{align}
\hat{\rho}_{[1,\ldots,L]}^{\rm erg}=&    \frac{1}{D_L}\sum_{\boldsymbol{s}\in \mathcal{B}_L}\ket{\boldsymbol{s}}_{[1,\ldots,L]}\bra{\boldsymbol{s}}= \\&   
\frac{1}{D_L}\biggl(\ket{0}_1\bra{0}\otimes\sum_{\boldsymbol{s}\in \mathcal{B}_{L-1}}\ket{\boldsymbol{s}}_{[2,\ldots,L]}\bra{\boldsymbol{s}} + \\&\ket{10}_{[1,2]}\bra{10}\otimes\sum_{\boldsymbol{s}\in \mathcal{B}_{L-2}}\ket{\boldsymbol{s}}_{[3,\ldots,L]}\bra{\boldsymbol{s}} \biggr);   
\end{align}
which, using $D_L = F_{L+2}$, can be rewritten as
\begin{align}
\hat{\rho}_{[1,\ldots,L]}^{\rm erg} =& \frac{F_{L+1}}{F_{L+2}} \ket{0}_1\bra{0}\otimes \hat{\rho}_{[2,\ldots,L]}^{\rm erg} +\\& \frac{F_{L}}{F_{L+2}} \ket{10}_{[1,2]}\bra{10}\otimes \hat{\rho}_{[3,\ldots,L]}^{\rm erg}.
\end{align}
By tracing out sites $[2,\ldots,L]$ we obtain the reduced state of the first site:
\begin{align}
\bigl(\hat{\rho}_{[1,\ldots,L]}^{\rm erg}\bigr)_1 \coloneqq &\operatorname{Tr}_{[2,\ldots,L]}\bigl[\hat{\rho}_{[1,\ldots,L]}^{\rm erg}\bigr]=\\&\frac{F_{L+1}}{F_{L+2}} \ket{0}\bra{0} + \frac{F_{L}}{F_{L+2}} \ket{1}\bra{1}.  
\end{align}
For simplicity, we denote the reduced state $\bigl(\hat{\rho}_{[1,\ldots,L]}^{\rm erg}\bigr)_j$ as $\hat{\rho}_j^{\rm erg}$.

The same reasoning can be applied recursively to the second site, yielding
\begin{widetext}
\begin{align}
\hat{\rho}_{[1,\ldots,L]}^{\rm erg}=&    
\frac{1}{D_L}\biggl[(\ket{0}_1\bra{0} + \ket{1}_1\bra{1})\otimes \ket{0}_2\bra{0}\otimes\sum_{\boldsymbol{s}\in \mathcal{B}_{L-2}}\ket{\boldsymbol{s}}_{[3,\ldots,L]}\bra{\boldsymbol{s}} + \ket{010}_{[1,2,3]}\bra{010}\otimes\sum_{\boldsymbol{s}\in \mathcal{B}_{L-3}}\ket{\boldsymbol{s}}_{[4,\ldots,L]}\bra{\boldsymbol{s}} \biggr]= \nonumber\\&
\frac{1}{F_{L+2}}\left[F_L(\ket{10}_{[1,2]}\bra{10}\otimes \hat{\rho}_{[3,\ldots,L]}^{\rm erg}+\ket{00}_{[1,2]}\bra{00}\otimes \hat{\rho}_{[3,\ldots,L]}^{\rm erg})+F_{L-1}(\ket{010}_{[1,2,3]}\bra{010}\otimes \hat{\rho}_{[4,\ldots,L]}^{\rm erg} \right],
\end{align}
\end{widetext}
and so
\begin{equation}
    \hat{\rho}_2^{\rm erg} = \frac{2 F_L}{F_{L+2}}\ket{0}\bra{0} + \frac{F_{L-1}}{F_{L+2}}\ket{1}\bra{1}.
\end{equation}
More generally, for a site $j$ in the bulk, the full ergodic state can be expressed as
\begin{widetext}
\begin{align}
    \hat{\rho}_{[1,\ldots,L]}^{\rm erg}=&\frac{1}{D_L}\Biggl[ \left(\sum_{\boldsymbol{s}\in \mathcal{B}_{j-1}}\ket{\boldsymbol{s}}_{[1,\ldots,j-1]}\bra{\boldsymbol{s}}\right) \otimes \ket{0}_j\bra{0}\otimes \left(\sum_{\boldsymbol{s}\in \mathcal{B}_{L-j}}\ket{\boldsymbol{s}}_{[j+1,\ldots,L]}\bra{\boldsymbol{s}}\right) +\nonumber\\& \left(\sum_{\boldsymbol{s}\in \mathcal{B}_{j-2}}\ket{\boldsymbol{s}}_{[1,\ldots,j-2]}\bra{\boldsymbol{s}}\right) \otimes \ket{010}_{[j-1,j,j+1]}\bra{010}\otimes \left(\sum_{\boldsymbol{s}\in \mathcal{B}_{L-j-1}}\ket{\boldsymbol{s}}_{[j+2,\ldots,L]}\bra{\boldsymbol{s}}\right) \Biggr].
\end{align}
\end{widetext}
Taking the partial trace over all sites except $j$ gives the general site-reduced ergodic state:
\begin{equation}
    \hat{\rho}_j^{\rm erg} = \frac{F_{j+1}F_{L-j+2}}{F_{L+2}}\ket{0}\bra{0} + \frac{F_{j}F_{L-j+1}}{F_{L+2}}\ket{1}\bra{1};
\end{equation}
Using $F_1 = F_2 = 1$ and $F_3 = 2$, one can verify that this expression holds for any site $j$ in the chain. This result implies that the ergodic expectation value of $\hat{Z}_j$ reads
\begin{equation}
    Z_j^{\rm erg}=\operatorname{Tr}\bigl[\hat{\rho}_j^{\rm erg}\hat{Z}_j\bigr]=\frac{F_{j}F_{L-j+1}-F_{j+1}F_{L-j+2}}{F_{L+2}}.
\end{equation}
From the diagonal structure of $\hat{\rho}_j^{\rm erg}$ it immediately follows that $X_j^{\rm erg}=Y_j^{\rm erg}=0$.

\section{Long time dynamics of $Y$}
\label{app:Ydynamics}
In this section, we report and briefly discuss the dynamics of the $Y$ operator. Following the discussion in the main text, Sec.~\ref{sec:thermalization}. Fig.~\ref{figapp:longtimeY} shows the dynamics for three different initial states (different panels) and three values of the driving field $h$ (different colors), with $\omega_d$ fixed.

The results reveal three distinct long-time dynamical regimes. For a system initialized in the polarized state, $Y$ weakly and noisily fluctuates around the ergodic expectation value, while for the Néel state, it exhibits more coherent oscillations. In contrast, the interpolating state $\ket{\Theta_+(\pi/4)}$ displays oscillations far from the ergodic value, with a strong dependence on $h$.

\end{document}